\documentclass[10pt]{article}
\usepackage{graphicx,times}
\usepackage{epsf}
\usepackage{amsmath,amssymb,latexsym,float,algorithm,algorithmic}
\usepackage{enumerate}
\usepackage{listings}
\numberwithin{equation}{section}
\newcommand{\norm}[1]{\left\Vert#1\right\Vert}

\newcommand{\Real}{\mathbb R}
\newcommand{\Complex}{\mathbb C}

\newcommand{\R}{\text{\fontshape{n}\selectfont I\kern-.42exR}}

\newcommand{\1}{\text{\fontshape{n}\selectfont 1\kern-.56exl}}

\begin{document}
\title{
{\bf QCDLAB:\\Designing Lattice QCD Algorithms with MATLAB}
}

\author{Artan Bori\c{c}i\\
        {\normalsize\it Physics Department, University of Tirana}\\
        {\normalsize\it Blvd. King Zog I, Tirana-Albania}\\
        {\normalsize\it borici@fshn.edu.al}\\
}

\date{}
\maketitle

\begin{abstract}
This paper introduces QCDLAB,
a design and research tool for lattice QCD algorithms.
The tool, a collection of MATLAB functions,
is based on a ``small-code'' and a ``minutes-run-time''
algorithmic design philosophy. The present version uses
the Schwinger model on the lattice,
a great simplification, which shares many
features and algorithms with lattice QCD.
A typical computing project using QCDLAB
is characterised by short codes, short run times,
and the ability to make substantial changes in a few seconds.
QCDLAB 1.0 can be downloaded from the QCDLAB project homepage
{\tt http://phys.fshn.edu.al/qcdlab.html}.
\end{abstract}

\pagebreak

\tableofcontents

\section{Advancement of Lattice QCD}

Lattice QCD, an industrial-range computing project, is in its fourth
decade. It has basically two major computing problems: simulation of
QCD path integral and calculation of quark propagators.
Generally, these problems lead to very intensive computations and require
high-end computing platforms.

However, we wish to make a clear distinction between lattice
QCD test and production codes. This is
very important in order to develop a compact and easily managable
computing project. While this is obvious in theory,
it is less so in lattice QCD practicing:
those who write lattice codes are focused primary on writing
production codes. What is usually called test code is merely
a test of production codes.

The code of a small project is  usually small,
runs fast, it is easy to access,
edit and debug. Can we achieve these features for a lattice test code?
Or, can we modify the goals of the lattice project in order to get such
features? In our opinion, this is possible for a {\it minimal test code,
a test code constisting of a minimal possible code
which is able to test gross features of the theory and algorithms at
shortest possible time and largest acceptable errors on a standard
computing platform}. This statement needs more explanation:

\begin{itemize}
\item[a.] Although it is hard to give sharp constraints
on the number of lines of the test code, we would call 
``minimal'' that code which is no more than a few
printed pages.
\item[b.] The run time depends on computing platforms and algorithms,
and the choice of lattice action and parameters.
It looks like a great number of degrees of
freedom here, but in fact there are hardly good choices in order to reduce
the run time of a test code
without giving up certain features of the theory.
Again, it is tremendously difficult to give run times.
However, a ``short'' run time should not
exceed a few minutes of wall-clock time.
\item[c.] We consider a computing platform as being ``standard''
if its cost is not too high for an academic computing project.
\item[d.] We call simulation errors to be the ``largest acceptable''
if we can distinguish clearly signal form noise
and when gross features of
the theory are not compromised by various approximations or choices.
\item[e.] Approximations should not alter basic features of the theory.
The quenched approximation, for example, should not be considered as an
acceptable approximation when studying QCD with light quarks.
\end{itemize}

A test code with these characteristics should
{\it signal the rapid advance in the field},
in which case, precision lattice computations
are likely to happen in many places around the world.
Writing a minimal test code is a {\it challenge of three smarts}:
smart computers, smart languages and smart algorithms.

\bigskip

In this paper we introduce the first version of QCDLAB, QCDLAB 1.0,
a collection of
MATLAB functions for the simulation of lattice Schwinger model.
This is part of a larger project for algorithmic development
in lattice QCD.
It can be used as a small laboratory to test and validate algorithms.
In particular, QCDLAB 1.0
serves as an illustration of the minimal test code concept.

QCDLAB can also be used for newcomers in the field. They can learn
and practice lattice projects which are based on short codes and run
times. This offers a ``learning by doing'' method, perhaps a quickest
route into answers of many unknown practical questions concerning
lattice QCD simulations.

The next two sections describe basic algorithms
for simulation of lattice QCD and foundations of Krylov subspace methods.
Then, we present the \mbox{QCDLAB 1.0} functions followed by examples
of simple computing projects. The last section outlines the future plans
of the QCDLAB project.

\section{Simulation of lattice QCD}

\subsection*{Notations and Problem definition}

Lattice gague fields $U_{\mu,i}$, our basic degrees of freedom,
are defined on oriented links \mbox{$i\rightarrow i+\hat\mu$}
of a four dimensional hypercubic lattice with $N$ sites
and lattice spacing $a$.
Here, $i$ is a four component index labeling the lattice sites,
$\mu=1,2,3,4$ labels dierctions in the Euclidean space,
and $\hat\mu$ is the unit vector along $\mu$-direction.
Algebraically, a lattice gauge field is an order 3, complex valued
unitary matrix with determinant one,
an element of the $SU(3)$ colour group.

The basic computational task in lattice QCD is the
generation of ensembles of guage field configurations
according to probability density:
$$
\rho_{QCD}(U) \sim \det (D^*D) ~e^{-S_g(U)}\ ,
$$
where $D$ is the lattice Dirac operator,
$$
S_g = \beta ~Re \sum_{i,\mu<\nu} (1 - P_{\mu\nu,i})
$$
is the gauge action, $\beta=6/g^2$ is the gauge-boson coupling constant, and
$$
P_{\mu\nu,i} = \frac13 
\text{tr} ~U_{\mu,i}U_{\nu,i+\hat\mu}U_{\mu,i+\hat\nu}^*U_{\nu,i}^*
$$
is the {\it plaquette} in $\mu\nu$ plane,
a 4-link product defined as in the figure.

\begin{picture}(100,180)(0,0)
\put(150, 40){\vector( 1, 0){100}}
\put(250, 40){\vector( 0, 1){100}}
\put(250,140){\vector(-1, 0){100}}
\put(150,140){\vector( 0,-1){100}}
\put(150, 30){$i$}
\put(250, 30){$i+\hat\mu$}
\put(250,140){$i+\hat\mu+\hat\nu$}
\put(120,140){$i+\hat\nu$}
\put(190, 20){$U_{\mu,i}$}
\put(260, 80){$U_{\nu,i+\hat\mu}$}
\put(190,150){$U_{\mu,i+\hat\nu}^*$}
\put(120, 80){$U_{\nu,i}^*$}
\end{picture}

We have assumed here a fermion theory with two degenerate flavours of
quark masses which suffices for the purpuse of this paper.
There are two main formulations of lattice fermions:
Wilson \cite{Wilson_1977} and Kogut-Susskind \cite{KS_1974}, 
called also staggered, discretizations of the Dirac operator.
Wilson operator, linking sites $i$ and $j$, is given by
$$
D_{ij}=(m+\frac 4a)I_4\otimes I_3~\delta_{ij}-\frac12\sum_{\mu}
[(I_4-\gamma_{\mu})\otimes U_{\mu,i}\delta_{i,j+\hat\mu}
+(I_4+\gamma_{\mu})\otimes U_{\mu,i-\hat\mu}\delta_{i,j-\hat\mu}]\ ,
$$
whereas Kogut-Susskind operator is given by
$$
D_{ij}=mI_3~\delta_{ij}+\frac12\sum_{\mu} (-1)^{i_1+\ldots+i_{\mu-1}}
(U_{\mu,i}\delta_{i,j+\hat\mu}-U_{\mu,i-\hat\mu}\delta_{i,j-\hat\mu})\ .
$$
Here $m$ is the bare quark mass, $\gamma_{\mu}$ are anticommuting
Hermitian {\it gamma-matrices} acting on Dirac space
$$
\gamma_{\mu}\gamma_{\nu}+\gamma_{\nu}\gamma_{\mu}=2\delta_{\mu\nu}I_4\ ,
$$
and $\otimes$ denotes the direct or Kronecker product of matrices.
Hence, Wilson and Kogut-Susskind operators are complex valued
matrices of order $12N$ and $3N$ with $49N$ and $25N$ nonzero
elements respectively.

Note that the difficulty of handling the determinat of a huge matrix
can be softened using the Gaussian integral expression:
\begin{equation}
\det (D^*D)
= \int \prod_{i=1}^N \frac{d\text{Re}(\phi_i) d\text{Im}(\phi_i)}{\pi}
~~e^{-\phi^* (D^*D)^{-1} \phi}
\end{equation}
where $\phi$ is a complex valued field,
a {\it pesudofermion} field. Thus, the determinant
is traded for the inversion. All we need now is to generate
ensembles according to the new density:
$$
\rho_{QCD}(U,\phi) \sim \exp\{-S_g(U)-\phi^*(DD^*)^{-1}\phi\}\ .
$$

\subsection*{Hybrid Monte Carlo Algorithm}

The HMC algorithm \cite{HMC_1987} starts by
introducing $su(3)$ conjugate momenta $P$ to $SU(3)$ lattice gauge fields.
Hence, the classical Hamiltonian can be written in the form
$$
\mathcal H(P,U,\phi) = \frac12\text{tr}P^2+S_g(U)+\phi^*(DD^*)^{-1}\phi\ ,
$$
whereas expanded probability density is
$$
\hat\rho_{QCD}(P,U,\phi) \sim \exp\{\mathcal H(P,U,\phi)\}\ .
$$

The idea of the HMC algorithm is as follows:
\begin{itemize}
\item[  i)] Use global heatbath for pseudofermion field update.
If $\zeta$ is a Gaussian pseudofermion field, the new field is
updated according to the equation
$$
\phi = D\zeta\ .
$$
\item[ ii)] Integrate numerically classical equations of motion.
\item[iii)] Correct numerical integration error using 
Metropolis {\it et al} algorithm.
\end{itemize}

\subsection*{Classical Equations of Motion}

The first first equation of motion is defined using conjugate
momenta:
$$
\dot U_{\mu,i} = iP_{\mu,i}U_{\mu,i}\ .
$$
For the second equation one writes down the total derivative
of the Hamiltonian:
$$
\mathcal {\dot H}=\sum_{\mu,i} \text{tr} P_{\mu,i} \dot P_{\mu,i}
+\dot S_g
-\phi^* (DD^*)^{-1} \dot DD^* (DD^*)^{-1}\phi + h.c.\ ,
$$
Substituting for $\dot U_{\mu,i}$ the first equation of motion and using
$$
\mathcal{\dot H}=0\ ,
$$
one finds the second equation of motion
$$
\dot P_{\mu,i}=F_{\mu,i}\ .
$$ 
From $\mathcal {\dot H}$ expression,
it is clear that in order to evaluate the force $F_{\mu,i}$,
one must calculate $(DD^*)^{-1}\phi$.

\subsection*{Leapfrog Algorithm}

The widely used algorithm for solving equations of motions is the
leapfrog algorithm
\begin{eqnarray*}
Q_{\mu,i}  &=& P_{\mu,i} + F_{\mu,i} \frac{\Delta t}2\\
U_{\mu,i}' &=& e^{iQ_{\mu,i}\Delta t} U_{\mu,i}\\
P_{\mu,i}' &=& Q_{\mu,i} + F_{\mu,i}'\frac{\Delta t}2\ ,
\end{eqnarray*}
where the primed fields are those advanced by $\Delta t$
and $Q_{\mu,i}$ are half-step momenta.
The algorithm starts at $t=0$, where
momenta are taken to be Gaussian noise. Then it continues
up to time $t=\tau$ for
$N_{\text{mic}}=\tau/\Delta t$ number of steps.

It is easy to show that this scheme is reversible and
preserves infinitesimal area of the phase space.
Reversibility guarantees detailed balance of HMC, whereas
area preservation ensures that there are no corrections
due to integration measure. However, Hamiltonian is not
conserved since
$$
H'-H=\frac12 \ddot{H} \Delta t^2 + O(\Delta t^3)\ .
$$

\subsection*{Metropolis {\it et al} Algorithm}

The HMC algorithm ends up accepting or rejecting the proposed
gauge field $\{U_{\mu,i}(\tau)\}$ using
Metropolis {\it et al} algorithm \cite{Metropolis_et_al_1953}.
The acceptance probability for this algorithm is
$$
P_{\text{acc}}(\{P(0),U(0)\}\rightarrow \{P(\tau),U(\tau)\})
=\min\left\{1,e^{H(\tau)-H(0)}\right\}
$$
On rejection, one goes back to time $t=0$ and refreshes momenta.

\subsection*{Inversion Algorithms} 

Calculation of forces and quark propagators requires the
solution of linear systems
$$
Dx=b\ ,
$$
where $D$ is a the lattice Dirac operator and $b$ the right hand side.
As we noted earlier, $D$ is a large and sparse matrix.
For these matrices,
{\it Krylov subspace methods} provide the most efficient
inversion algorithms \cite{my_thesis_1996}.
Such an algorithm is the Conjugate Gradients (CG) algorithm
\cite{CG_1952}.

\subsection*{Conjugate Gradients}

Given an approximation $x_0$, the algorithm starts with computation of the
{\it residual vector} $r_0$,
$$
r_0=b-Ax_0\ ,
$$
and initialisation of a vector $p_0$ to the starting residual, $p_0=r_0$.
Note that CG assumes that $A$ is a {\it positive definite} and Hermitian
matrix. Then, the algorithm iterates these vectors using recursions
\begin{eqnarray*}
  x_{k+1}&=&x_k+\alpha_k p_k\\
  r_{k+1}&=&r_k-\alpha_k Ap_k\\
  p_{k+1}&=&r_{k+1}+\beta_{k+1} p_k\ ,
\end{eqnarray*}
where
$$
\alpha_k=\frac{r_k^*r_k}{p_k^*Ap_k}, ~~~~~~~~
\beta_{k+1}=\frac{r_{k+1}^*r_{k+1}}{r_k^*r_k}\ .
$$

\subsection*{Conjugate Gradients on Normal Equations}

Since the lattice Dirac operator is neither positive definite nor Hermitian
one takes $A=D^*D$ and solves for the {\it normal equations}
$$
D^*Dx=D^*b\ .
$$
Then, we get what we call the Conjugate Gradients algorithm on Normal Equations
(CGNE). As with standard CG, given an approximation $x_0$,
the algorithm computes starting residual $r_0$,
$$
r_0=b-Dx_0\ ,
$$
and initialises $p_0$ to $r_0$. Additionally,
a new vector $s_0$ is initialised using $s_0=D^*r_0$.
The CGNE algorithm recursions are
\begin{eqnarray*}
  x_{k+1}&=&x_k+\alpha_k p_k\\
  r_{k+1}&=&r_k-\alpha_k Dp_k\\
  p_{k+1}&=&s_{k+1}+\beta_{k+1} p_k\ ,
\end{eqnarray*}
where
$$
s_{k+1}=D^*r_{k+1}\ ,
$$
and
$$
\alpha_k=\frac{s_k^*s_k}{(Dp_k)^*(Dp_k)}, ~~~~~~~~
\beta_{k+1}=\frac{s_{k+1}^*s_{k+1}}{s_k^*s_k}\ .
$$

\subsection*{A Convergence Result}

It can be theoretically proven
that CGNE algorithm converges {\it linearly}, i.e.
$$
\norm{r_k} \leq 2\left(\frac{\kappa(D)-1}{\kappa(D)+1}\right)\norm{r_{k-1}}\ ,
$$
where $\norm{.}$ denotes the Euclidean norm and $\kappa(D)$ is the
{\it condition number} of the Dirac operator. In terms of
{\it singluar values} of $D$, $\sigma_1\geq\sigma_2\geq\cdots\geq\sigma_N$,
the condition number is given by
$$
\kappa(D)=\frac{\sigma_1}{\sigma_N}\ .
$$
If $D$ is rescaled such that $\sigma_1=1$ we get
(see \cite{my_thesis_1996}, p52)
$$
\norm{r_k} \leq 2\left(\frac{1-\sigma_N}{1+\sigma_N}\right)^k\norm{r_0}\ .
$$
It can be shown that CGNE calculates a solution of the minimisation problem
$$
\min_{x\in \Complex^N}\norm{b-Dx}\ .
$$
But for non-normal matrices, such as the Wilson-Dirac operator,
this solution is {\it sub-optimal}.
Unlike CGNE, the General Minimised Residual (GMRES) algorithm
calculates the {\it optimal} solution of this problem.
Both methods convegre according to the law described above.
However, $\sigma_N$ is traded for the spectral gap, which is larger
(for non-normal matrices).
As we will show later, GMRES algorithm is more expensive and often
prohibitive in terms of computer resources.

\section{Foundations of Krylov Subspace Methods}

A Krylov subspace is the space built from the pair $(r_0,D)$:
$$
{\mathcal K}_k = span\{r_0,Dr_0,\ldots,D^{k-1}r_0\}\ ,
$$
where $1\leq k\leq m$, with $m$ being the rank of $D$.
The simpliest example consists of the pair $r_0$ and the
identity matrix $I$, in which case $k=1$ and the
Krylov subspace is simply the vector $r_0$. In this case we have to do with an
{\it invariant subspace}: further multiplications of $r_0$ by $D=I$
will not increase the subspace.

Iterative methods which seek solutions $x$ in ${\mathcal K}_k$
are called Krylov subspace methods. If $Q_k=[q_1,\ldots,q_k]$ is
a basis of othonormal vectors of ${\mathcal K}_k$,
the approximate solution can be written as
$$
x_k=x_0+ \sum_{i=1}^k y_i q_i
$$
In order to compute it one has to compute first $q_i,i=1,2,\ldots,k$. 
There are two general approaches to compute $y$, namely
\begin{itemize}
\item[~i)] Galerkin approach: choose $y$ such that the residual vector $r_k$ is orthogonal to ${\mathcal K}_k$.
\item[ii)] Minimal residual approach: choose $y$ such that
$\norm{r_k}$ is minimal.
\end{itemize}

\subsection*{Basis generation: Arnoldi Algorithm}

The method is a modified Gram-Schmidt orthogonalisation of ${\mathcal K}_k$,
in which the next vector is computed by
$$
{\tilde q}_{k+1} = D q_k - \sum_{j=1}^k q_j h_{jk}\ ,
$$
and where the coefficients $h_{jk}$ are chosen such that
the vectors come mutually orthonormal:
$$
q_j^*{\tilde q}_{k+1}=0.
$$
From this condition we get
$$
h_{jk}=q_j^*Dq_k\ .
$$
The algorithm that facilitates this process is called
{\it Arnoldi algorithm} \cite{Arnoldi_1951}.
\begin{algorithm}[htp]
\caption{\hspace{.2cm} Arnoldi algorithm}
\label{Arnoldi_alg}
\begin{algorithmic}
\STATE $\rho=\norm{r_0}$
\STATE $q_1 = r_0 / \rho$
\FOR{$~k = 1, \ldots$}
    \STATE $w = D q_k$
    \FOR{$~j = 1, \ldots, k$}
       \STATE $h_{kj} = q_j^* w$
       \STATE $w := w - q_j h_{jk}$
    \ENDFOR
    \STATE $h_{k+1,k} = \norm{w}$
    \IF{$h_{k+1,k} = 0$} \STATE stop \ENDIF
    \STATE $q_{k+1} = w / h_{k+1,k}$
\ENDFOR
\end{algorithmic}
\end{algorithm}
Having basis vectors one can construct two linear solvers,
which are described in the following.

\subsection*{FOM: Full Orthogonalisation Method}

If we denote $H_k$ the matrix with elements $h_{ij},i,j=1,\ldots,k$,
the result of Arnoldi decomposition can be written in matrix form:
$$
D Q_k = Q_k H_k + h_{k+1,k}  q_{k+1} e_k^T \equiv Q_{k+1} {\tilde H}_k\ .
$$
Approximate solution can also be written as
$$
x_k=x_0+Q_ky_k\ .
$$
For the residual error vector one can write:
\begin{eqnarray*}
r_k&=&b-Dx_k\\
   &=&r_0-DQ_ky_k\\
   &=&q_1\rho - Q_kH_ky_k-h_{k+1,k}q_{k+1}e_k^Ty_k\ .
\end{eqnarray*}
The Galerkin approach requires the next residual to be orthonormal to all
previous vectors
$$
Q_k^*r_k=0\ ,
$$
which is
$$
Q_k^*q_1\rho-Q_k^*Q_kH_ky_k-Q_k^*q_{k+1}h_{k+1,k}e_k^Ty_k= 0\ .
$$
Using orthonormality of $Q_k$,
\begin{eqnarray*}
Q_k^*q_1&=&e_1,\\
Q_k^*Q_k&=&I_k,\\
Q_k^*q_{k+1}&=&0\ ,
\end{eqnarray*}
one obtains the linear system
$$
H_ky_k=e_1\rho\ .
$$
Note that $H_k$ is an upper Hessenberg matrix and that the size of
the problem depends on the value of $k$, which is usually a much smaller
than $N$, the order of the original problem.

\subsection*{GMRES: Generalised Minimal Residual Method}

Arnoldi recurrences can be written also in the form:
$$
D Q_k = Q_k H_k + h_{k+1,k}  q_{k+1} e_k^T \equiv Q_{k+1} {\tilde H}_k\ ,
$$
where now ${\tilde H}_k$ is an upper Hessenberg $(k+1)\times k$ matrix,
or the matrix $H_k$ appended by the row $h_{k+1,k}e_k^T$. With this
notations the residual error vector can be written as
$$
r_k=q_1\rho - Q_{k+1}{\tilde H}_k y_k\ .
$$
The minimal residual strategy of GMRES \cite{GMRES_1986}
requires that
$$
\norm{b-Dx_k} \rightarrow \min, ~~~x_k \in {\cal K}_k\ .
$$
Substituting $x_k=Q_ky_k$ and $q_1=Q_{k+1}e_1$ we get
$$
\norm{Q_{k+1}(e_1\rho-{\tilde H}_k y_k)} \rightarrow \min\ .
$$
Since $Q_{k+1}$ is orthonormal, it can be ignored and we get the
smaller least squares problem:
$$
\norm{e_1\rho-{\tilde H}_k y_k} \rightarrow \min, ~~~y_k \in \Complex^k\ .
$$

\subsection*{Krylov solvers are Polynomial Approximation solvers}

In fact, the approximate solution in the Krylov subspace
$$
x_k \in {\mathcal K}_k = span\{r_0,Dr_0,\ldots,D^{k-1}r_0\}\ ,
$$
can be described as a degree $k-1$ polynomial applied to $r_0$:
$$
x_k=x_0+P_{k-1}(D)r_0\ .
$$
Then, the residual vector is a degree $k$ polynomial
applied to $r_0$:
\begin{eqnarray*}
r_k&=&r_0-DP_{k-1}(D)r_0\\
&=&\left[I-DP_{k-1}(D)\right]r_0\\
&\equiv& R_k(D)r_0\ .
\end{eqnarray*}
Hence, GMRES solves the constrained minimisation problem:
find the polynomial $R_k$ such that
$$
\norm{R_k(D)r_0} \rightarrow \min\ , ~~~~~R_k(0)=1\ .
$$
In fact, this is a characterisation of Krylov subspace methods.
One speaks of {\it optimal polynomials} generated in this way.

\subsection*{But ...}

... GMRES requires to store all Arnoldi vectors and its
work grows proportionally to $k^2$. One can limit this growth of resources
by restarting the algorithm after a given number of steps. Using this strategy,
robustness is lost and sometimes convergence as well.
Going back to normal equations,
$$
D^*Dx=D^*b\ ,
$$
we know that the optimal polynomial is computed for $D^*D$ and {\it not}
for $D$ itself. However, computing resources remain constant in this case,
an important advantage over GMRES. Therefore, a great deal of
research has been devoted to methods which are as cheap as CGNE and yet
have similar convergence to GMRES.

\subsection*{BiCG$\gamma_5$}

One of these methods is the
specialisation of the Biconjugate Gradients (BiCG) algorithm in the case
of the Wilson operator, which is $\gamma_5$-Hermitian
(see \cite{my_thesis_1996} p47):
$$
D^*=\gamma_5 D\gamma_5\ .
$$
The method, coined BiCG$\gamma_5$, can be formally obtained from CG
by inserting a $\gamma_5$ operator whenever a scalar product occurs:
$$
u^*v\longrightarrow u^*\gamma_5 v\ .
$$
Since, $\gamma_5$ is a nondefinite operator, this scalar product may not
exist, and a premature breakdown may occur. In practice we see an
irregular behaviour of the residual vector norm history. 

\subsection*{BiCGStab}

Biconjugate Grandients Stabilised algorithm, or BiCGStab
\cite{BiCGStab_1992}
replaces the redundant recursion of BiCG for a local minimiser of the
residual vector norm,
thus giving a general and robust solver for non-Hermitian systems.

\section{QCDLAB 1.0}

QCDLAB is designed to be a high level language interface
for lattice QCD computational procedures. 
It is based on the MATLAB and OCTAVE language and environment.
While MATLAB is a product of The MathWorks, OCTAVE is its clone,
a free software under the terms of the GNU General Public License.

MATLAB/OCTAVE is a technical computing environment integrating numerical
computation and graphics in one place, where problems and solutions look
very similar and sometimes almost the same as they are written mathematically.
Main features of MATLAB/OCTAVE are:

\begin{itemize}
\item Vast Build-in mathematical and linear algebra functions.
\item Many functions form Blas, Lapack, Minpack, etc. libraries.
\item State-of-the-art algorithms.
\item Interpreted language.
\item Dynamically loaded modules from other languages like C/C++, FORTRAN.
\item Ability to compile OCTAVE codes using the Startego Octave
Compiler, {\tt Octave-Compiler.org}.
\end{itemize}

Hence, QCDLAB offers a two level language system: a higher level language,
which is very popular for numerical work and a lower level translation to
C++. In fact, if required the lower level can be further optimised for
the particular hardware in place.

The first version of QCDLAB is intended for work on the higher level
only. QCDLAB 1.0 contains the following MATLAB/OCTAVE functions:

\vspace{0.5cm}
{\tt
\begin{tabular}{lllll}
Autocorel  & BiCGg5    & BiCGstab    & Binning   & cdot5\\
CG         & CGNE      & Dirac\_KS   & Dirac\_r  & Dirac\_W\\
FOM        & Force\_KS & Force\_W    & GMRES     & HMC\_KS\\
HMC\_W     & Lanczos   & SCG         & SUMR      & wloop
\end{tabular}
}
\vspace{0.5cm}

We divide them in two groups: simulation and inversion algorithms.
We begin below with the description of simulation algorithms.

\subsection{Simulation Algorithms}

This section introduces QCDLAB 1.0 simulation tools of lattice QED2.
In this case, lattice gauge fields $U_{\mu,i}$ 
can be expressed using angles $\theta_{\mu,i}\in\Real$,
$$
U_{\mu,i}=e^{i\theta_{\mu,i}}\ ,
$$
whereas gauge action is given by
$$
S_g = \beta \sum_{i,\mu<\nu} [1-
\cos(\theta_{\mu,i}+\theta_{\nu,i+\hat\mu}-\theta_{\mu,i+\hat\nu}-\theta_{\nu,i})]\ ,
$$
where $\beta=1/e^2$, and electron charge $e$ .

\subsection*{Dirac Operators}

In case of Wilson fermions the Dirac operator is given by
$$
D_{ij}=(m+\frac 4a)I_2~\delta_{ij}-\frac12\sum_{\mu=1}^2
[(I_2-\sigma_{\mu})U_{\mu,i}\delta_{i,j+\hat\mu}
+(I_2+\sigma_{\mu})U_{\mu,i-\hat\mu}\delta_{i,j-\hat\mu}]\ ,
$$
with $\sigma_{\mu}$ being Pauli matrices,
$$
\sigma_1=
\begin{pmatrix}
0&1\\
1&0
\end{pmatrix},~~~~
\sigma_2=
\begin{pmatrix}
0&-i\\
i&0
\end{pmatrix},~~~~
\sigma_3=
\begin{pmatrix}
1&0\\
0&-1
\end{pmatrix}\ .
$$
QCDLAB defines spin projection operators in terms of these matrices
$$ 
{\mathcal P}_{\mu}^{+}=\frac12(I_2+\sigma_{\mu}), ~~~~
{\mathcal P}_{\mu}^{-}=\frac12(I_2-\sigma_{\mu})\ .
$$
They are computed using this code:
\begin{small}
\begin{lstlisting}
% Form spin projection operators
P1_plus = [1,1;1,1]/2; P1_minus = [1,-1;-1,1]/2;
P2_plus = [1,-i;i,1]/2; P2_minus = [1,i;-i,1]/2;
\end{lstlisting}
\end{small}

Given the quark mass, {\tt mass},
the number of lattice sites along each direction, {\tt N},
the total number of lattice sites, {\tt N2},
the gauge field configuaration, {\tt U1}, and spin projector operators
as input, {\tt Dirac\_W} returns Wilson matrix, {\tt A1}:
\begin{small}
\begin{lstlisting}
A1=Dirac_W(mass,N,N2,U1,P1_plus,P2_plus,P1_minus,P2_minus);
\end{lstlisting}
\end{small}

For staggered fermions we have
$$
D_{ij}=m~\delta_{ij}+\frac12\sum_{\mu=1}^2 \epsilon_{\mu,i}
(U_{\mu,i}\delta_{i,j+\hat\mu}-U_{\mu,i-\hat\mu}\delta_{i,j-\hat\mu})\ ,
$$
where
$$
\epsilon_{1,i}=1, ~~~~~~\epsilon_{2,i}=(-1)^{i_1}\ .
$$
{\tt Dirac\_KS} below returns the staggered matrix:
\begin{small}
\begin{lstlisting}
A1=Dirac_KS(mass,N,N2,U1);
\end{lstlisting}
\end{small}

\subsection*{Forces}

In order to compute the force it is convenient to have ready a
nearest neighbours list for each lattice site. The code that implement
forward {\tt kp} and backward {\tt km} lists is given below.
\begin{small}
\begin{lstlisting}
% Make nearest neighbours list
for j2=0:N-1;
for j1=0:N-1;
  k       = 1 +            j1 +            j2*N;
  kp(k,1) = 1 + mod(j1+1  ,N) +            j2*N;
  kp(k,2) = 1 +            j1 + mod(j2+1  ,N)*N;
  km(k,1) = 1 + mod(j1-1+N,N) +            j2*N;
  km(k,2) = 1 +            j1 + mod(j2-1+N,N)*N;
end
end
\end{lstlisting}
\end{small}

In case of two degenerate Wilson fermions, the force is given by
$$
F_{\mu,i}=-\beta\sin(\theta_{\mu,i}+\theta_{\nu,i+\hat\mu}
-\theta_{\mu,i+\hat\nu}-\theta_{\nu,i})
+2Re ~iU_{\mu,i}^* (\chi^*_{i+\hat\mu}\mathcal P_{\mu}^+\eta_i
    +\eta^*_{i+\hat\mu}\mathcal P_{\mu}^-\chi_i)\ ,
$$
where pseudofermion fields are two-component complex valued vectors.
{\tt Force\_W} returns both gauge {\tt pg} and fermion {\tt pf} pieces.
Its arguments are:
number of lattice sites, {\tt N2}, forward neighbour list, {\tt kp},
backward neighbour list, {\tt km}, angles, {\tt theta1},
pseudofermion fields, {\tt eta}, {\tt chi}, gauge field, {\tt U1},
and spin projector operators.
\begin{small}
\begin{lstlisting}
[pf,pg]=Force_W(N2,kp,km,theta1,eta,chi,U1,P1_plus,P2_plus,P1_minus,P2_minus);
\end{lstlisting}
\end{small}

In case of four degenerate Kogut-Susskind fermions the force is
$$
F_{\mu,i}=-\beta
\sin(\theta_{\mu,i}+\theta_{\nu,i+\hat\mu}-\theta_{\mu,i+\hat\nu}-\theta_{\nu,i})
+2Re ~iU_{\mu,i}^*(\epsilon_{\mu,i+\hat\mu}\chi^*_{i+\hat\mu}\eta_i
    -\epsilon_{\mu,i}\eta^*_{i+\hat\mu}\chi_i)\ ,
$$
where pseudofermion fields are complex values numbers. As in Wilson case,
{\tt Force\_KS} returns gauge {\tt pg} and fermion {\tt pf} forces.
\begin{small}
\begin{lstlisting}
[pf,pg]=Force_KS(N2,kp,km,theta1,eta,chi,U1);
\end{lstlisting}
\end{small}

\subsection*{Simulation Tools}

QCDLAB's simulation tools are {\tt HMC\_W} and {\tt HMC\_KS}.
Their arguments are:
\begin{itemize}
\item[] {\tt iconf}: set to zero for {\it hot} start,
\item[] {\tt theta1}: starting angles.
\end{itemize}
On completion they return:
\begin{itemize}
\item[] {\tt A2}: Dirac operator on {\tt theta2} background,
\item[] {\tt Plaq}: plaquette history,
\item[] {\tt Q\_top}: topological charge history,
\item[] {\tt Wloop}: ten smallest Wilson loops history,
\item[] {\tt theta2}: output angles,
\item[] {\tt stat}: a four column array of Metropolis test history.
\end{itemize}

One trajectory of Hybrid Monte Carlo algorithm consists of steps
given in Algorithm \ref{hmc_traj_algor}.
\begin{algorithm}
\caption{\hspace{.2cm} HMC Trajectory in QCDLAB}
\label{hmc_traj_algor}
\begin{itemize}
\item Heatbath update of pseudofermion fields {\tt phi=A1*eta0;}
\item Heatbath update of momenta {\tt P=randn(N2,2);}
\item Compute Hamiltonian {\tt H1}.
\item Invert Dirac operator {\tt chi=A1'\verb+\+eta0;}
\item Advance momenta half step {\tt P=P+pdot*deltat/2;}
\item Start molecular dynamics loop.
\begin{itemize}
\item Advance angles full step {\tt theta2=theta2+P*deltat;}
\item Compute inversions {\tt eta=A2\verb+\+phi; chi=A2'\verb+\+eta;}
\item Advance momenta full step {\tt P=P+pdot*deltat;}
\end{itemize}
\item Advance angles full step {\tt theta2=theta2+P*deltat;} 
\item Compute inversions {\tt eta=A2\verb+\+phi; chi=A2'\verb+\+eta;}
\item Advance momenta half step {\tt P=P+pdot*deltat/2;}
\item Compute Hamiltonian {\tt H2}.
\item Perform Metropolis test.
\end{itemize}
\end{algorithm}
In case of a successful test, the function computes
topological charge,
$$
Q_{\text{top}} = \frac1{2\pi}\sum_{i,\mu<\nu}
\sin(\theta_{\mu,i}+\theta_{\nu,i+\hat\mu}-\theta_{\mu,i+\hat\nu}-\theta_{\nu,i})\ ,
$$
and $n\times n$ Wilson loops,
$$
W_n = \frac1N\sum_{i,\mu<\nu}
\cos\left(
 \sum_{k=0}^{n-1} \theta_{\mu,i+k\hat\mu}
+\sum_{k=0}^{n-1} \theta_{\nu,i+n\hat\mu+k\hat\nu}
-\sum_{k=0}^{n-1} \theta_{\mu,i+n\hat\nu+k\hat\mu}
-\sum_{k=0}^{n-1} \theta_{\nu,i+k\hat\nu}\right)\ .
$$
{\tt HMC\_W} and {\tt HMC\_KS} functions are called in the form given below.

\begin{small}
\begin{lstlisting}
[A2,Plaq, Q_top, Wloop, theta2, stat] = HMC_W(theta1,iconf);
[A2,Plaq, Q_top, Wloop, theta2, stat] = HMC_KS(theta1,iconf);
\end{lstlisting}
\end{small}

\subsection*{Computation of Wilson Loops}

{\tt wloop} returns ten smallest Wilson loops. 
Angles $\theta_{\mu,i},\theta_{\mu,i+\hat\mu},
\ldots,\theta_{\mu,i+9\hat\mu}$ are implemented using
arrays {\tt tleg1},{\tt tleg2}...{\tt tlog10}.
These are summed over to give arrays {\tt wleg1},{\tt wleg2}...{\tt wlog10}.
Then, each of these arrays is used to compute the corresponding Wilson loop
around a square. The function is called using:
\begin{small}
\begin{lstlisting}
wlp = wloop(N,N2,kp,km,theta1);
\end{lstlisting}
\end{small}

\subsection*{Autocorrelations and Errors}

When measuring an observable, such as plaquette, we get time series
of data in the form ${\cal O}_i, i=1,2,\ldots,n$. Standard error
estimation procedures rely on the assumption that data are
{\it decorrelated}. This can be
checked by measuring the autocorrelation function
$$
A(t)=\text{cov}({\cal O}(0){\cal O}(t)) \sim e^{-t/t_{\text{exp}}}\ ,
$$
where $t_{\text{exp}}$ is called {\it exponential autocorrelation time}.
{\tt Autocorel}, which implements $A(t)$, has two arguments:
the data vector, {\tt x}, and the maximal time interval, {\tt t}:
\begin{small}
\begin{lstlisting}
y=Autocorel(x,t);
\end{lstlisting}
\end{small}

In fact, a quick way to estimate the error is to block or `bin' data.
In this case one computes block averages and estimates the error
of averages for increasing block size. {\tt Binning} does exactly that.
One must specify the original data, {\tt x}, and the maximal block size,
{\tt t}. It returns a $t$-element vector {\tt err}, containing error
estimates for block sizes $1,2,\ldots,t$:
\begin{small}
\begin{lstlisting}
err=Binning(x,t);
\end{lstlisting}
\end{small}
A simple recipe is to take $t\sim t_{\text{exp}}$ and choose
the maximum value of {\tt err}.

\subsection*{Computing Projects}

The best way to test QCDLAB capabilities
is to set up a simple computing project
like the following:
{\it Compute square Wilson loops and topological charge using {\tt HMC\_KS}.
Graph Wilson loops as a function of the linear size.
Do you get a perimeter law? Plot the histogram of the topological charge.
How is it distributed?}

As usuall, one opens two windows, one running MATLAB/OCTAVE,
and one text editor where {\tt HMC\_KS.m} file is located.
We present here an exmaple of running OCTAVE with model and
algorithmic parameters as in the listing. Entering
\begin{lstlisting}
[A2,Plaq,Q_top,Wloop,theta2,stat]=HMC_KS([],0);
\end{lstlisting}
one gets an output stream that looks something like:
\begin{lstlisting}
ans = 
   3.00000   0.33333   0.57985  -1.25858 
ans = 
   16.00000    0.12500    0.76736   -0.92516 
ans = 
   20.00000    0.15000    0.75921   -0.27965
\end{lstlisting}
We have displayed here only the first three lines.
Columns of the stream display trajectory number, acceptance,
plaquette and topological charge.
One can store the angle output, {\tt theta2}, and feed it into the next run:
\begin{lstlisting}
theta1=theta2;
[A2,Plaq,Q_top,Wloop,theta2,stat]=HMC_KS(theta1,1);
\end{lstlisting}
In our example project we ran seven batches of {\tt HMC\_KS} and
analysed results of the last batch.
We computed and plotted autocorelation functions of five Wilson loops:
\begin{lstlisting}
auto=Autocorel(Wloop(:,1),20);
semilogy(auto)
hold
auto=Autocorel(Wloop(:,2),20);
semilogy(auto)
auto=Autocorel(Wloop(:,3),20);
semilogy(auto)
auto=Autocorel(Wloop(:,4),20);
semilogy(auto)
auto=Autocorel(Wloop(:,5),20);
semilogy(auto)
xlabel('t')
ylabel('Autocorel')
replot
gset terminal postscript
gset out 'Auto.ps'
replot
gset terminal x11
hold
\end{lstlisting}
\centerline{\includegraphics[scale=.5,angle=-90]{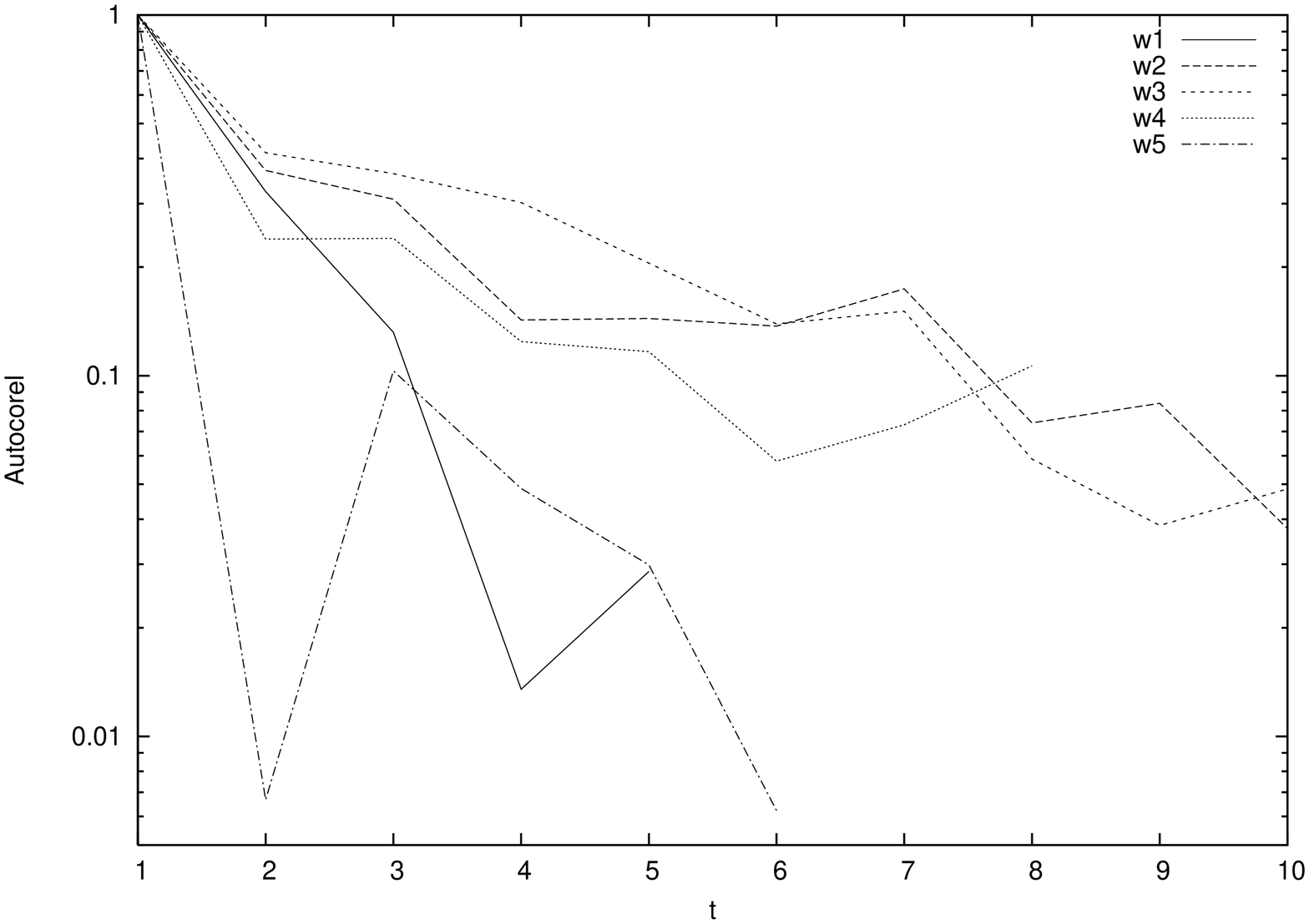}}

Then we computed central values of Wilson loops:
\begin{lstlisting}
w=mean(Wloop);
\end{lstlisting}
The errors are estimated using {\tt Binning} with
the largest block size set to 10:
\begin{lstlisting}
sw=max(Binning(Wloop(:,1),10));
sw=[sw,max(Binning(Wloop(:,2),10))];
sw=[sw,max(Binning(Wloop(:,3),10))];
sw=[sw,max(Binning(Wloop(:,4),10))];
sw=[sw,max(Binning(Wloop(:,5),10))];
sw=[sw,max(Binning(Wloop(:,6),10))];
sw=[sw,max(Binning(Wloop(:,7),10))];
sw=[sw,max(Binning(Wloop(:,8),10))];
sw=[sw,max(Binning(Wloop(:,9),10))];
sw=[sw,max(Binning(Wloop(:,10),10))];
\end{lstlisting}
Then, to plot Wilson loops we entered:
\begin{lstlisting}
semilogyerr(w(1:5),sw(1:5))
hold
axis([0,6,1e-4,1])
semilogy(w(1:5))
axis([0,6,1e-3,1])
xlabel('Linear Size')
ylabel('Wloop')
replot
gset terminal postscript
gset out 'Wloop.ps'
replot
gset terminal x11
hold
\end{lstlisting}

\centerline{\includegraphics[scale=.5,angle=-90]{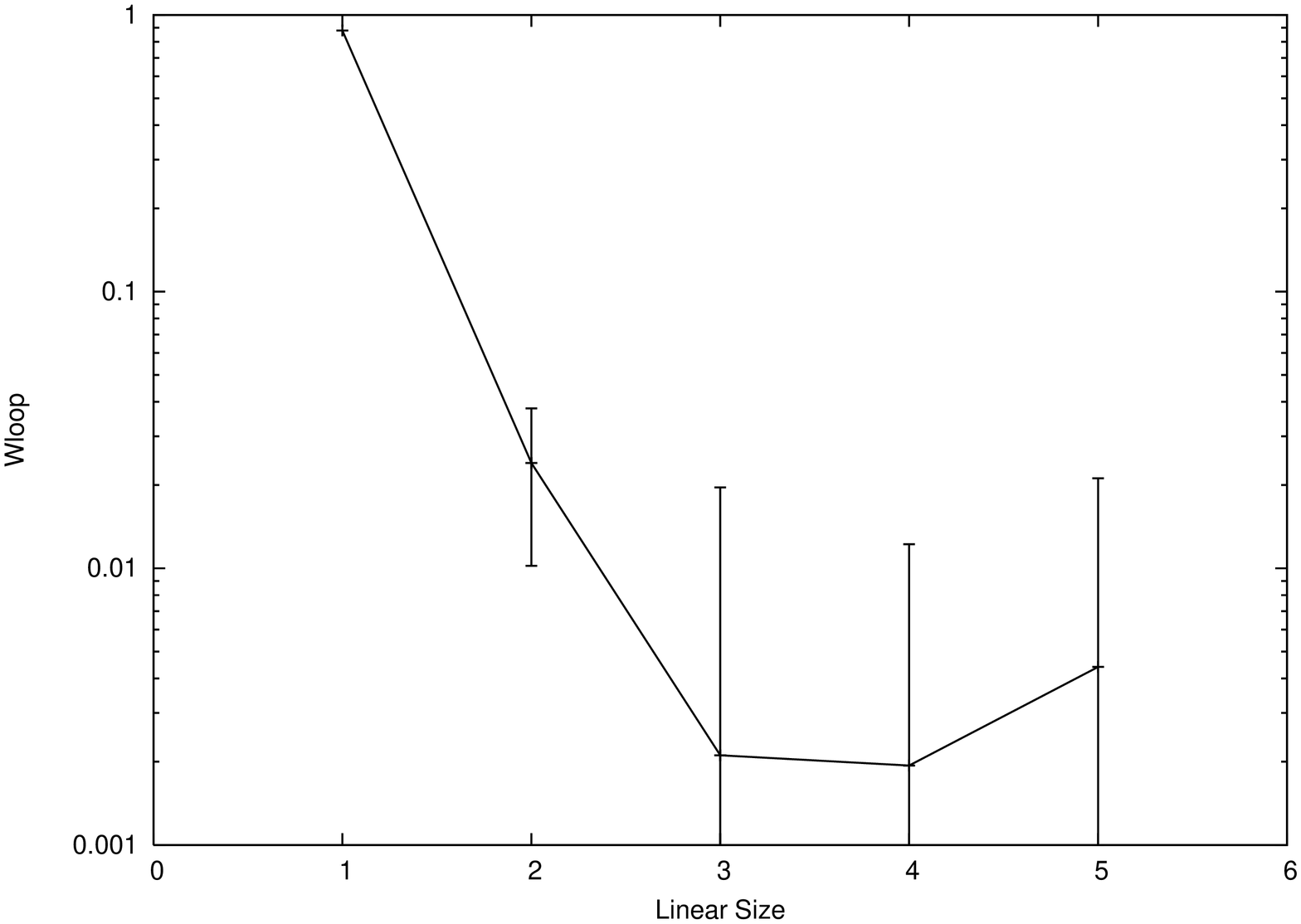}}

Finally, to produce the histogram plot of topological charge is very easy:
\begin{lstlisting}
hist(Q_top,30)
xlabel('Q_top')
ylabel('Frequency')
gset terminal postscript
gset out 'Q_top.ps'
replot
gset terminal x11
hold
\end{lstlisting}

\centerline{\includegraphics[scale=.5,angle=-90]{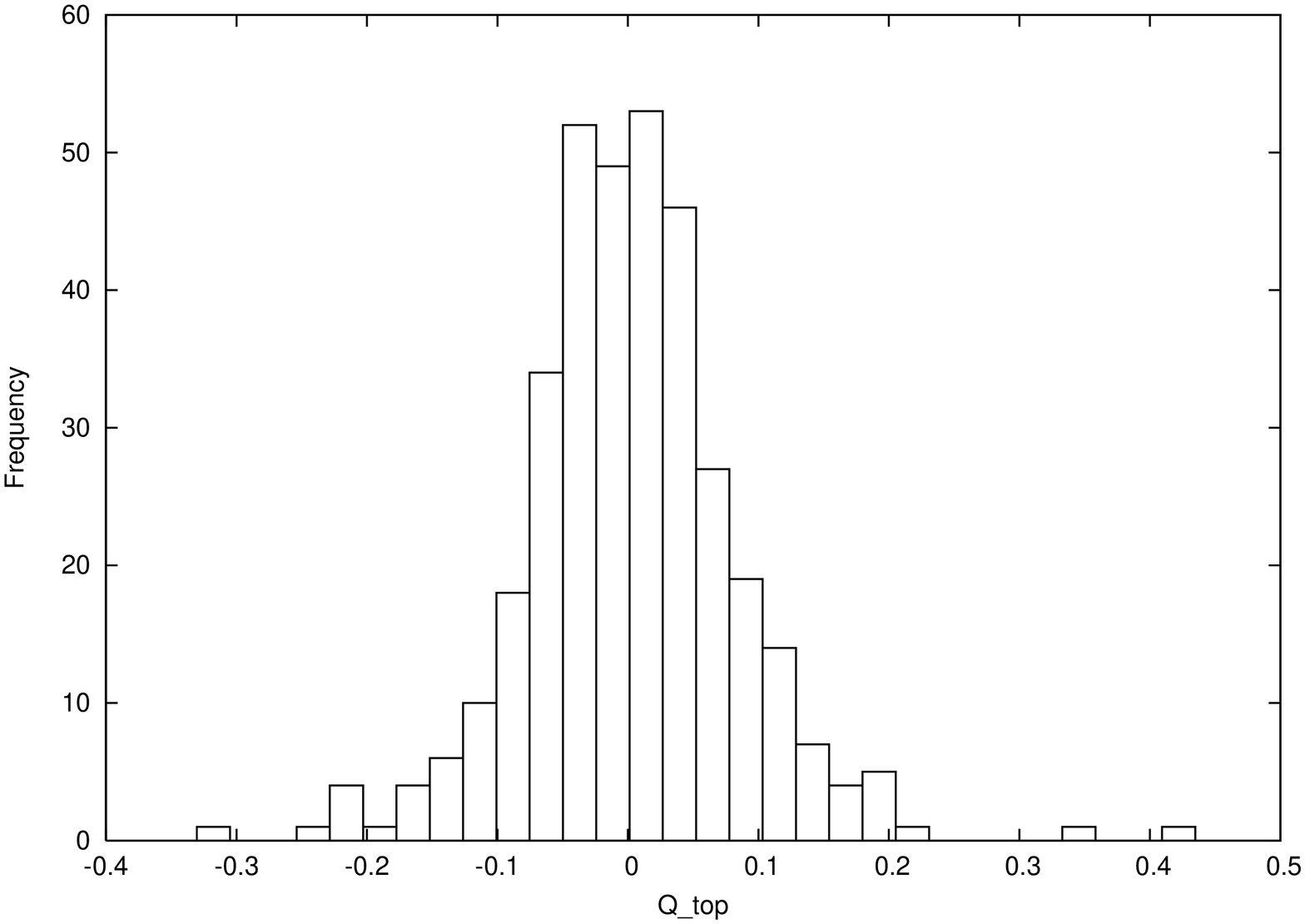}}

\subsection{Inversion Algorithms}

In this section we list QCDLAB functions which implement basic
Krylov subspace inverters for use in lattice guage theories.
In this version the whole matrix {\tt A}, be it sparse or dense,
should be supplied as an argument, together with the right
hand side {\tt b}, the approximate soluction {\tt x\_0},
the tolerance {\tt tol} and the maximum number of iterations {\tt nmax}.
{\tt CG.m} and {\tt CGNE.m} functions below return solution {\tt x} and
{\it recursive} residual vector norm history {\tt rr}.

\begin{small}
\begin{lstlisting}
[x,rr] = CG(A,b,x0,tol,nmax);
[x,rr] = CGNE(A,b,x0,tol,nmax);
\end{lstlisting}
\end{small}

{\tt FOM.m} and {\tt GMRES.m} return {\it true} residual
instead of recursive residual and Arnoldi matrix {\tt H} as well.
This can be used to compute approximate egienvalues of the
original matrix $A$.
\begin{small}
\begin{lstlisting}
[z,rt,H] = FOM(A,b,z0,tol,nmax);
[z,rt,H] = GMRES(A,b,z0,tol,nmax);
\end{lstlisting}
\end{small}

\subsection*{BiCG$\gamma_5$ and BiCGstab}

The {\tt BiCGg5} function
may be used for $\gamma_5$-Hermitian operators:
\begin{small}
\begin{lstlisting}
[x,rr] = BiCGg5(A,b,x0,tol,nmax);
\end{lstlisting}
\end{small}
It uses a special inner product,
a pseudo-scalar which does not lead
to a vector norm. Hence, the algorithm can break down prematurely.
There are {\it look ahead} strategies which cure this problem.
We don't employ them. However, to avoid a starting breakdown
the recipe is to use a non-trivial initial solution.

BiCGstab is a general non-Hermitian solver.
It inherits from BiCG the premature breakdown problem.
In order to avoid a starting breakdown we use a random
initial {\it left Lanczos vector} {\tt y0}.
\begin{small}
\begin{lstlisting}
[z,rr]=BiCGstab(A,b,z0,tol,nmax);
\end{lstlisting}
\end{small}

\subsection*{Symmetric Lanczos algorithm}

The counterpart of Arnoldi algorithm for Hermitian matrices is
Lanczos algorithm \cite{Lanczos_1952}. From Arnoldi algorithm we have
$$
H_k = Q_k^*AQ_k\ .
$$
If $A$ is Hermitian we can write
$$
H_k^*= Q_k^*A^*Q_k = Q_k^*AQ_k = H_k\ .
$$
Since $H_k$ is upper Hessenberg and Hermitian,
it must be tridiagonal. It is commonly denoted by $T_k$.
This way, after $k$ Lanczos steps we have
$$
A Q_k = Q_k T_k + \beta_k  q_{k+1} e_k^T\ ,
$$
where
$$
T_k=
\begin{pmatrix}
 \alpha_1 & \beta_1     &        & \\
 \beta_1  & \alpha_2    & \ddots & \\
          & \ddots      & \ddots       & \beta_{k-1} \\
          &             & \beta_{k-1}  & \alpha_k \\
\end{pmatrix}
$$

QCDLAB function {\tt Lanczos} arguments are matrix {\tt A},
starting vector {\tt b} and maximal number of steps {\tt nmax}.
It returns Lanczos vectors {\tt Q} and matrix {\tt T}:
\begin{small}
\begin{lstlisting}
[Q,T]=Lanczos(A,b,nmax);
\end{lstlisting}
\end{small}

\subsection*{Computing Projects}

In this section we illustrate QCDLAB inverter functions for
Wilson fermions. The matrix $A$, a function argument,
is an output of {\tt HMC\_W}. Having an angle configuration,
one can generate it using {\tt Dirac\_r} function:
given the quark mass {\tt mass}, the Wilson parameter {\tt r},
the lattice size {\tt N}, the total number of lattice sites {\tt N2},
and the angle configuration {\tt theta1} as arguments,
it returns the Wilson-Dirac matrix {\tt A1}:
\begin{small}
\begin{lstlisting}
A1=Dirac_r(mass,r,N,N2,theta1);
\end{lstlisting}
\end{small}

Note that knowing $A$ is not essential.
Indeed, any user supplied procedure of matrix-vector multiplication
can be called whenever {\tt =A*} occurs in the function.
Future releases of QCDLAB will provide capabilities that implement this.

\subsection*{Wilson Fermions}

In our example project here, we used an angle configuration on
a $16\times 16$ lattice. We loaded this configuration and
created a right hand side and a starting solution. Then we generated
three Wilson-Dirac matrices of three different fermion masses:
\begin{lstlisting}
load theta16
N=16;N2=N^2;b=zeros(2*N2,1);x0=b;b(1)=1;
A1=Dirac_r(-0.1,1,N,N2,theta2);
A2=Dirac_r(-0.05,1,N,N2,theta2);
A3=Dirac_r(0,1,N,N2,theta2);
\end{lstlisting}

A first thing to do is to
compute and plot eigenvalues of the massless operator:
\begin{lstlisting}
e3=eig(A3);
plot(e3,'o')
\end{lstlisting}

\centerline{\includegraphics[scale=.6,angle=-90]{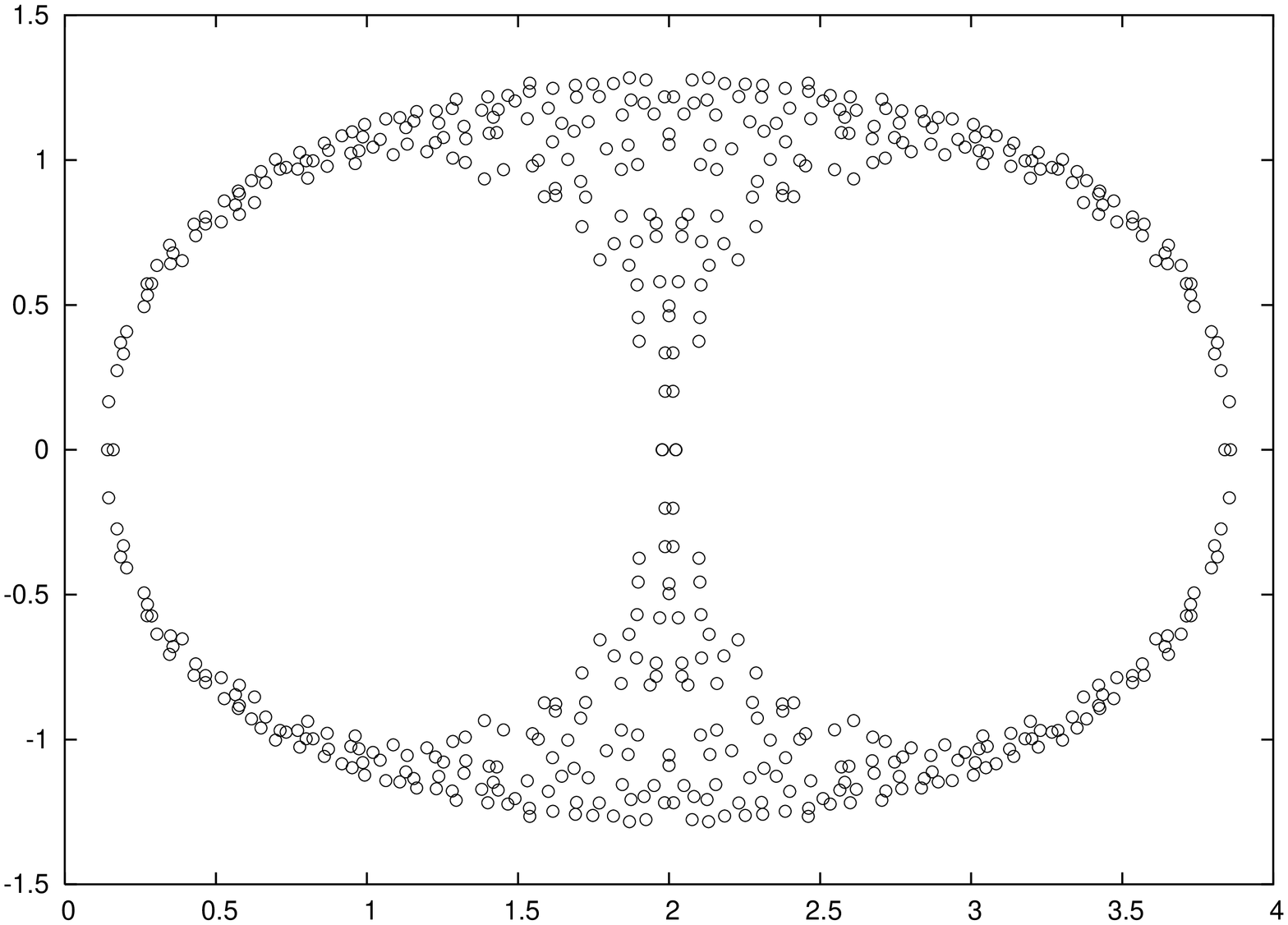}}

In order to compare GMRES and CGNE convergence one calls
respective solvers for each fermion mass as follows:
\begin{lstlisting}
tol=1e-13;nmax=2*N2;
[x_gmres1,r_gmres1,H1] = GMRES(A1,b,x0,tol,nmax);
[x_gmres2,r_gmres2,H2] = GMRES(A2,b,x0,tol,nmax);
[x_gmres3,r_gmres3,H3] = GMRES(A3,b,x0,tol,nmax);
[x_cgne1,r_cgne1] = CGNE(A1,b,x0,tol,nmax);
[x_cgne2,r_cgne2] = CGNE(A2,b,x0,tol,nmax);
[x_cgne3,r_cgne3] = CGNE(A3,b,x0,tol,nmax);
\end{lstlisting}
Then one plots the residual norm history as a function of
matrix-vector multiplications calls:
\begin{lstlisting}
k_gmres1=max(size(r_gmres1));
k_gmres2=max(size(r_gmres2));
k_gmres3=max(size(r_gmres3));
k_cgne1=2*max(size(r_cgne1));
k_cgne2=2*max(size(r_cgne2));
k_cgne3=2*max(size(r_cgne3));
semilogy(1:k_gmres1,r_gmres1,';gmres1;')
hold
semilogy(1:k_gmres2,r_gmres2,';gmres2;')
semilogy(1:k_gmres3,r_gmres3,';gmres3;')
semilogy(1:2:k_cgne1,r_cgne1,';cgne1;')
semilogy(1:2:k_cgne2,r_cgne2,';cgne2;')
semilogy(1:2:k_cgne3,r_cgne3,';cgne3;')
xlabel('# matrix-vector')
ylabel('residual norm')
replot
gset terminal postscript
gset out 'conv_hist.ps'
replot
gset terminal x11
hold
\end{lstlisting}
Note that GMRES has an additional overhead which grows like $i^2$,
rendering the algorithm useless for large $i$. Hence, the GMRES
convergence, measured in terms of matrix-multiplication number,
should be considered as a theoretically ideal result and a
benchmark for the performance of short-recurrences algorithms.

\centerline{\includegraphics[scale=.6,angle=-90]{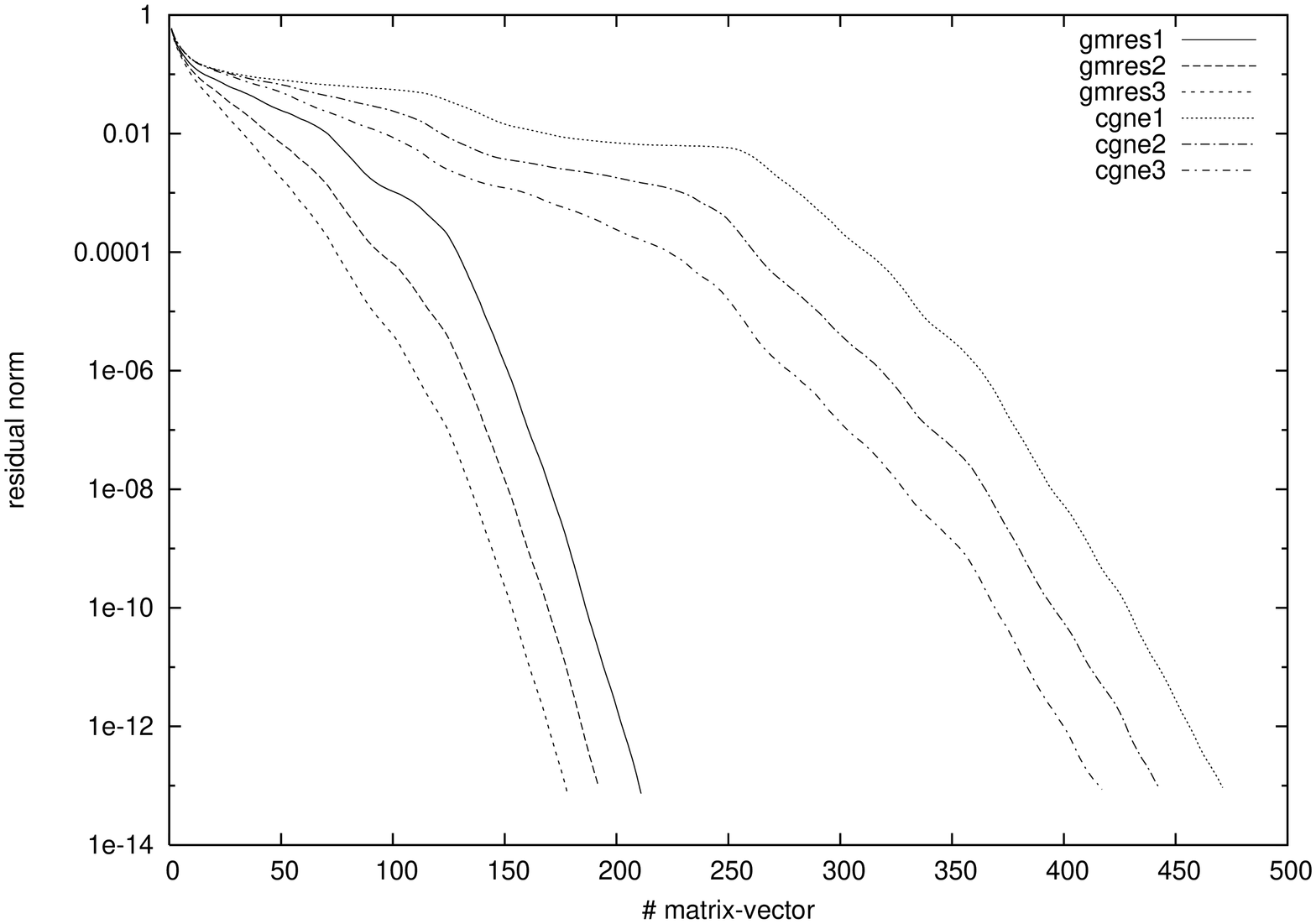}}

The difficulty of GMRES exploding resources can be avoided
using {\tt BiCGg5} and {\tt BiCGstab} functions.
Employing the lightest mass one can compare convergence history of all
solvers:
\begin{lstlisting}
x0=rand(2*N2,1);
[x_gmres1,r_gmres1,H1] = GMRES(A1,b,x0,tol,nmax);
[x_cgne1,r_cgne1] = CGNE(A1,b,x0,tol,nmax);
[x_bicg1,r_bicg1] = BiCGg5(A1,b,x0,tol,nmax);
[x_bicgstab1,r_bicgstab1]=BiCGstab(A1,b,x0,tol,nmax);
k_gmres1=max(size(r_gmres1));
k_cgne1=2*max(size(r_cgne1));
k_bicg1=max(size(r_bicg1));
k_bicgstab1=2*max(size(r_bicgstab1));
semilogy(1:k_gmres1,r_gmres1,';gmres1;')
hold
semilogy(1:2:k_cgne1,r_cgne1,';cgne1;')
semilogy(1:k_bicg1,r_bicg1,';bicg1;')
semilogy(1:2:k_bicgstab1,r_bicgstab1,';bicgstab1;')
xlabel('# matrix-vector')
ylabel('residual norm')
replot
gset terminal postscript
gset out 'all_solver_conv_hist.ps'
replot
gset terminal x11
hold
\end{lstlisting}

\centerline{\includegraphics[scale=.6,angle=-90]{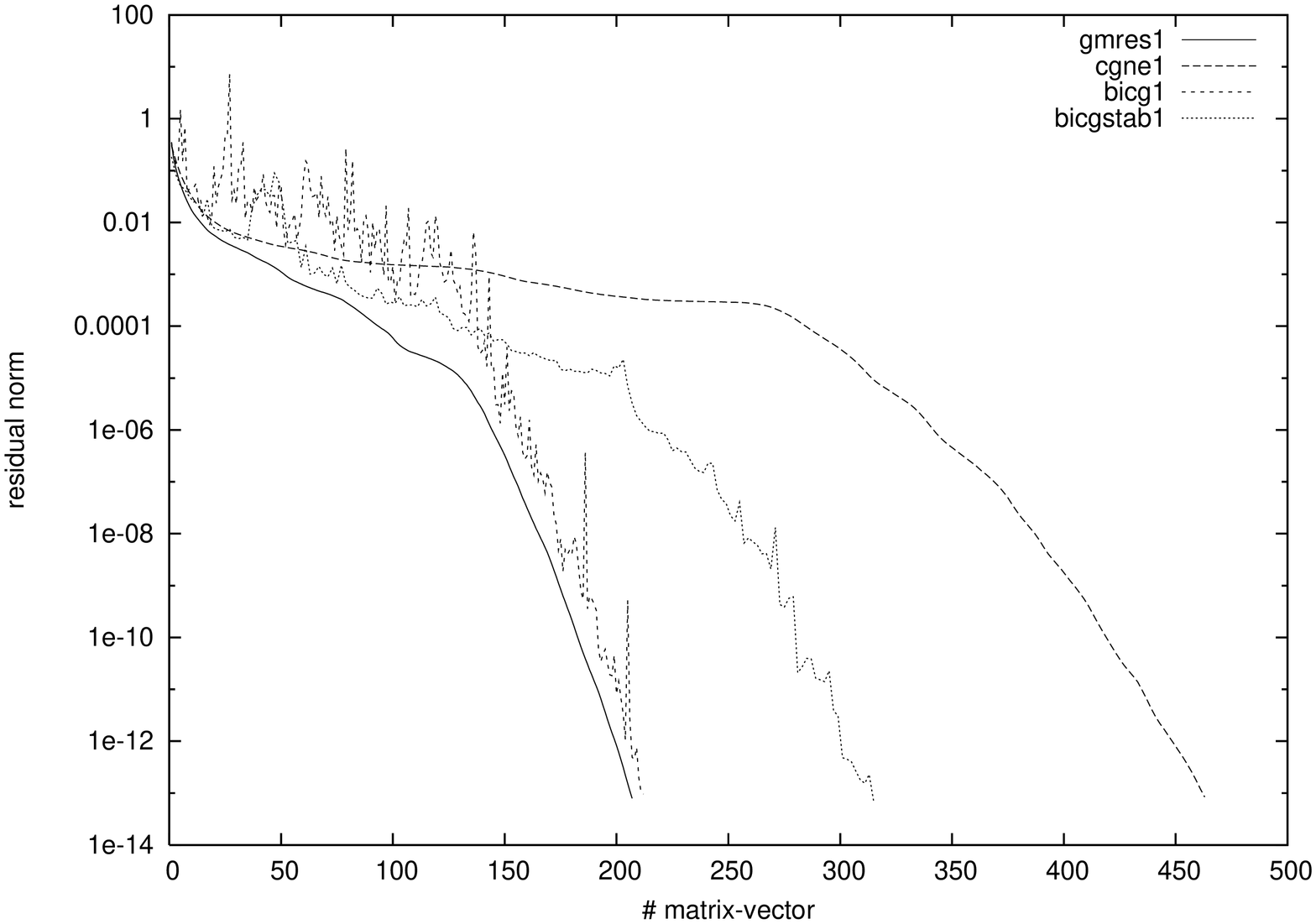}}

As seen from the plot, the best short-recurrences
solver (for this particular example) is
{\tt BiCGg5}. It converges at about the same matrix-vector multiplications
as {\tt GMRES} and yet avoiding its pitfalls.
Its irregular convergence history can be softened to the level of
{\tt BiCGstab} using the {\it quasiminimal residual approach},
or QMR algorithm, which is not described here.

\subsection*{Staggered Fermions}

Staggered operator is anti-Hermitian as can be illustrated below:
first create the staggered matrix; then typing {\tt norm(A+A')}, the answer
will be zero. Since {\tt i*A} is Hermitian, its eigenvalues will be
real. We plotted the eigenvalues as sorted from the {\tt eig} function.
\begin{lstlisting}
U1=cos(theta2)+sqrt(-1)*sin(theta2);
N=16;N2=N^2;
A=Dirac_KS(0,N,N2,U1);
norm(A+A')
ea=eig(i*A);
plot(ea,'o;ea;')
\end{lstlisting}

Theory tells that for normal matrices, such as staggered operator,
CGNE is an optimal solver. The following plot compares GMRES and CGNE
convergence history as a function of matrix-vector multiplications
counter.
\begin{lstlisting}
b=zeros(N2,1);x0=b;b(1)=1;
tol=1e-13;nmax=N2;
[x_gmres,r_gmres,H] = GMRES(A,b,x0,tol,nmax);
[x_cgne,r_cgne] = CGNE(A,b,x0,tol,nmax);
k_cgne=max(size(r_cgne));
semilogy(r_gmres,';gmres;')
hold
semilogy(1:2:2*k_cgne,r_cgne,';cgne;')
xlabel('# matrix-vector')
ylabel('residual norm')
replot
gset terminal postscript
gset out 'conv_hist_ks.ps'
replot
gset terminal x11
hold
\end{lstlisting}

\hspace{0mm}\centerline{\includegraphics[scale=.56,angle=-90]{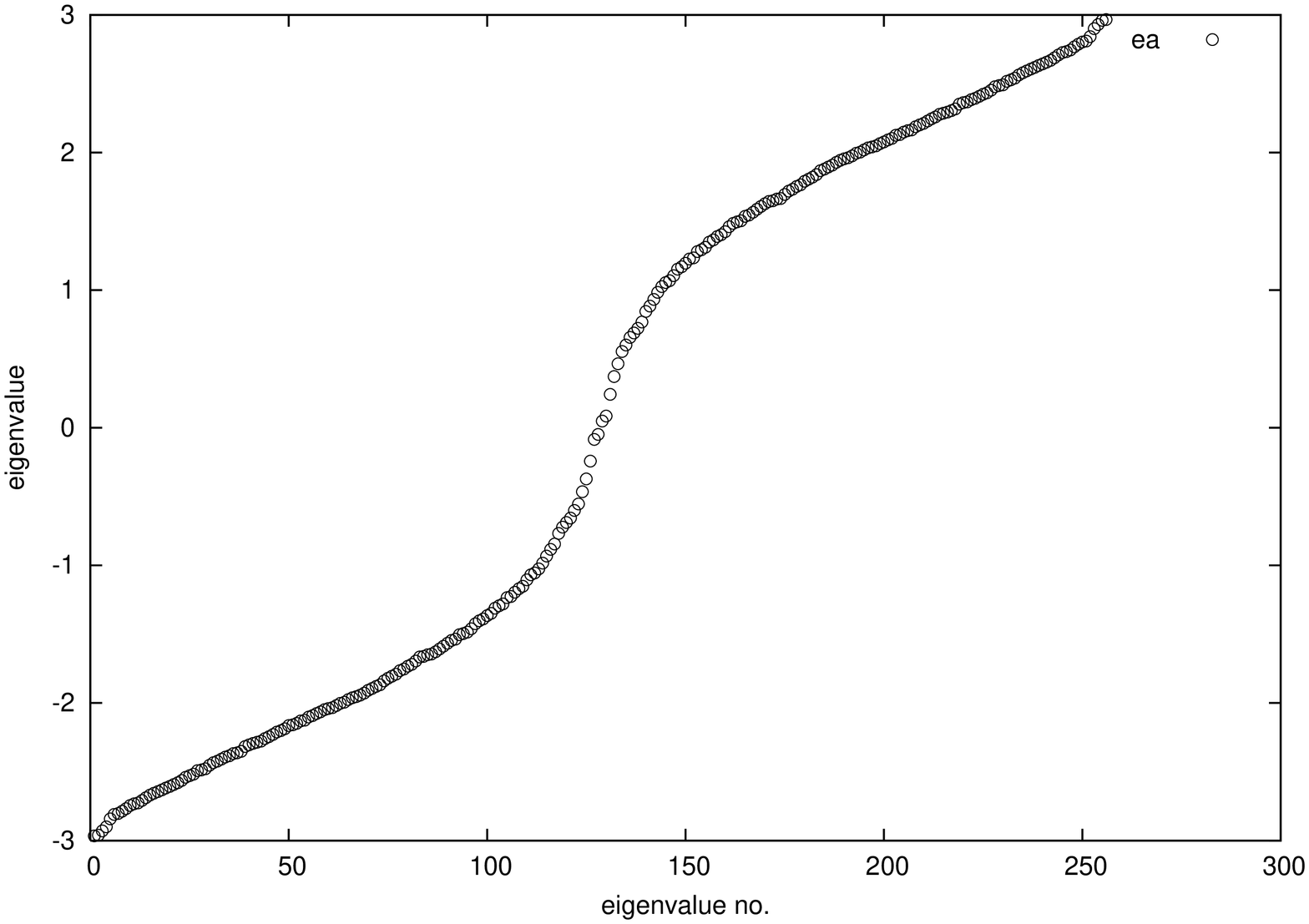}}

\vspace{2cm}

\centerline{\includegraphics[scale=.6,angle=-90]{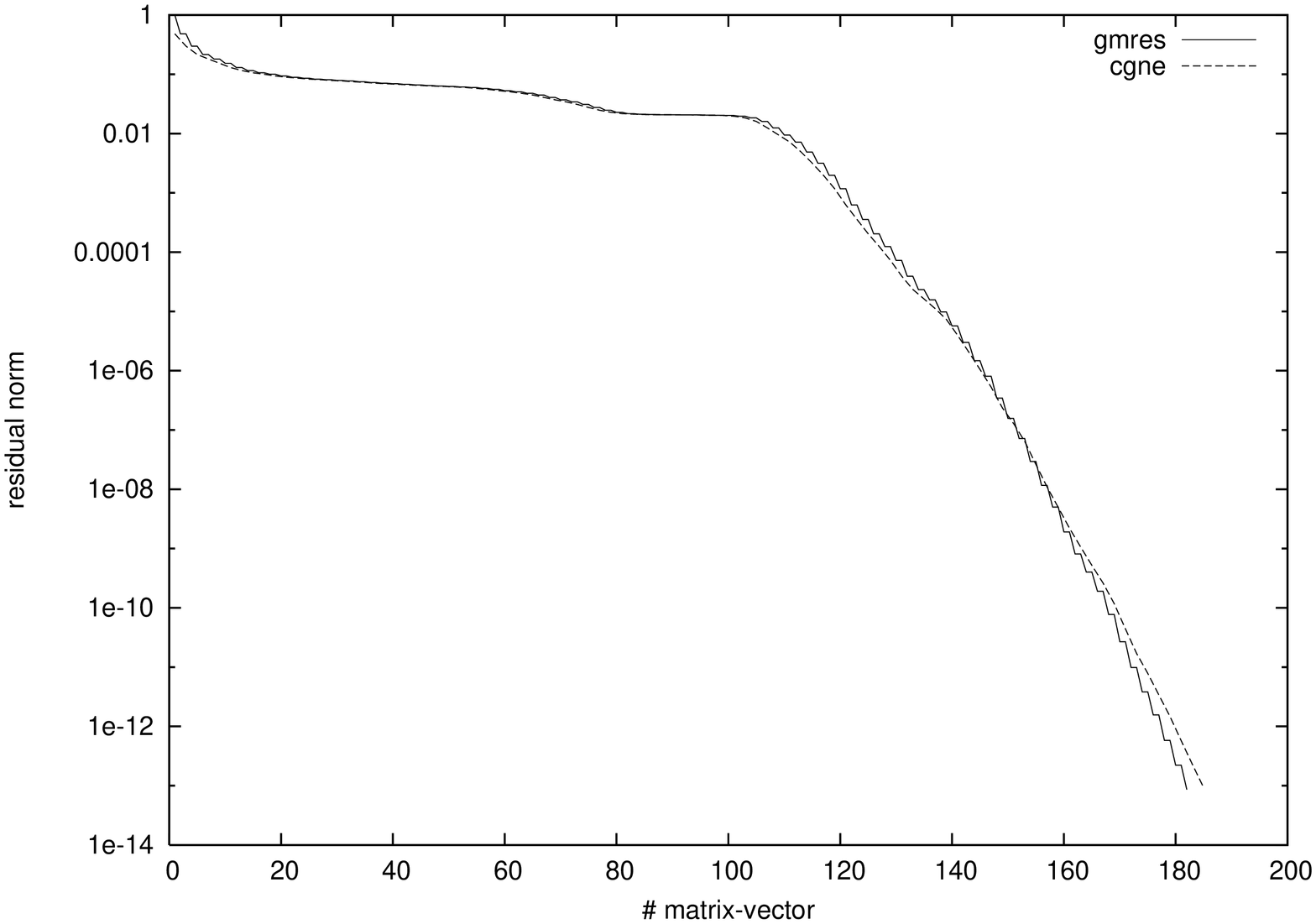}}

\subsection*{CG-Lanczos Equivalence}

As another application, one can use Lanczos algorithm to solve
the linear system
$$
D^*Dx=D^*b
$$
for Wilson matrix $D$. Taking the lightest mass from the previous exmaple
and setting maximal iteration number to {\tt k\_cgne3}:
\begin{lstlisting}
[Q,T]=Lanczos(A3'*A3,A3'*b,k_cgne3);
\end{lstlisting}
one can construct the solution using:
\begin{lstlisting}
x_lanczos3=Q(:,1:k_cgne3)*(T\[norm(A3'*b);zeros(k_cgne3-1,1)]);
\end{lstlisting}
Comparison to {\tt x\_cgne3} yields:
\begin{lstlisting}
norm(x_cgne3-x_lanczos3)
ans =  2.5537e-14
\end{lstlisting}
This example illustrates the theoretical result that CG and Lanczos
algorithms are equivalent linear solvers in exact arithmetic.

\subsection{Ginsparg-Wilson Fermions}

In this section we describe QCDLAB functions for use with lattice chiral
fermions. A chiral lattice Dirac operator satisfies the Gisnparg-Wilson
relation \cite{Ginsparg_Wilson_1982}:
\begin{equation*}
\gamma_5 D + D\gamma_5 = 2D\gamma_5D\ .
\end{equation*}
One solution to this relation is the Nueberger overlap operator
\cite{Neuberger_1998},
\begin{equation*}
D=\frac{1+m}2 ~I + \frac{1-m}2 \gamma_5 \text{sgn}(H_W)\ ,
\end{equation*}
where $sgn(.)$ is the signum function, $H_W=\gamma5 D_W$.
For the signum function to be nontrivial, the Wilson-Dirac operator should be
indefinite, which is the case if its bare mass $M$ is sufficiently negative.
This is usually taken to be in the interval $(-2,0)$.

Another form of the Neuberger operator is
\begin{equation*}
D=\frac{1+m}2 ~I + \frac{1-m}2 D_W \left(D_W^*D_W\right)^{-\frac12}\ .
\end{equation*}
If we express $D_W$ in terms of its singular values and vectors,
\begin{equation*}
D_W=U\Sigma V^*\ ,
\end{equation*}
we get
\begin{equation*}
D=\frac{1+m}2 ~I + \frac{1-m}2 UV^*\ ,
\end{equation*}
Since $U$ and $V$ are unitary operators, so it is $UV^*$. Hence, the
overlap operator is a shifted unitary operator. Iterative inverters for
such operators can be simplified as we show below. Before doing this
we give an iterative method in order to compute the overlap operator.

\subsection*{Lanczos Algorithm for the Overlap Computation}

In order to apply the overlap operator to a vector $b$ one should first
perform the inversion $(D_W^*D_W)^{1/2} x = b$ and then apply
$D_W$ to x.
The calculation is based on the following integral representation
for the inverse square root \cite{Borici_isqr_2000}:
\begin{equation*}
(D_W^*D_W)^{-1/2} = \frac{2}{\pi} \int_0^{\infty} dt (t^2 + D_W^*D_W)^{-1}\ .
\end{equation*}
From the previous section we know that one can use the Lanczos algorithm
to solve the linear systems, such as
\begin{equation}
(D_W^*D_W)^{-1} b = Q_k T_k^{-1} e_1 \rho\ ,
\end{equation}
where $\rho = \norm{b}$.
Since by shifting the matrix $D_W^*D_W$ one obtains the same Lanczos
vectors, we can write
\begin{equation}
(t^2 + D_W^*D_W)^{-1} b = Q_k (t^2 + T_k)^{-1} e_1 \rho\ .
\end{equation}
Using the above integral representation again, but now for Lanczos matrix
$T_k$ we get
\begin{equation}\label{result}
x = (D_W^*D_W)^{-1/2} b = Q_k T_k^{-1/2} e_1 \rho\ .
\end{equation}
To summarise, in order to find $x$ one computes:
\begin{itemize}
\item $Q_k$ and $T_k$ using the {\tt Lanczos} function on $D_W^*D_W$ and $b$,
\item then computes $y_k = T_k^{-1/2} e_1 \rho$,
\item and finally $x = Q_k y_k$.
\end{itemize}
Using the {\tt A3,b} pair of the last example one can enter these commands:
\begin{lstlisting}
b=rand(2*N2,1);
rho=norm(b);
[Q,T]=Lanczos((A3-eye(512))'*(A3-eye(512)),b,200);
[N,k]=size(Q);
e1=zeros(k-1,1); e1(1)=1;
y=sqrtm(T)\e1*rho;
x=(A3-eye(512))*Q(:,1:k-1)*y;
\end{lstlisting}
In case $Q_k$ vectors are too large to be stored in the main memory of the
computer, one can modify {\tt Lanczos} such that it does not accumulate
Lanczos vectors into Q.
In this case, one calculates $y_k$ and then repeats the Lanczos iteration
in order to form $x$. This is the so called {\it double pass algorithm}.

For small problems, as it is usually the case for QED2 on the lattice, one
can compute the overlap operator using the singular value decomposition:
\begin{lstlisting}
[U_W,S_W,V_W]=svd(A3-eye(512));
V=U_W*V_W';
z=V*b;
norm(z-x)
D=eye(512)+V;
eigD=eig(D);
plot(eigD,'+')
\end{lstlisting}
In this case it is easy to compute $D$ eigenvalues using direct methods,
such as {\tt eig} function. The following plot shows the eignevalues of
the massless overlap operator $D$.

\hspace{0mm}\centerline{\includegraphics[scale=.56,angle=-90]{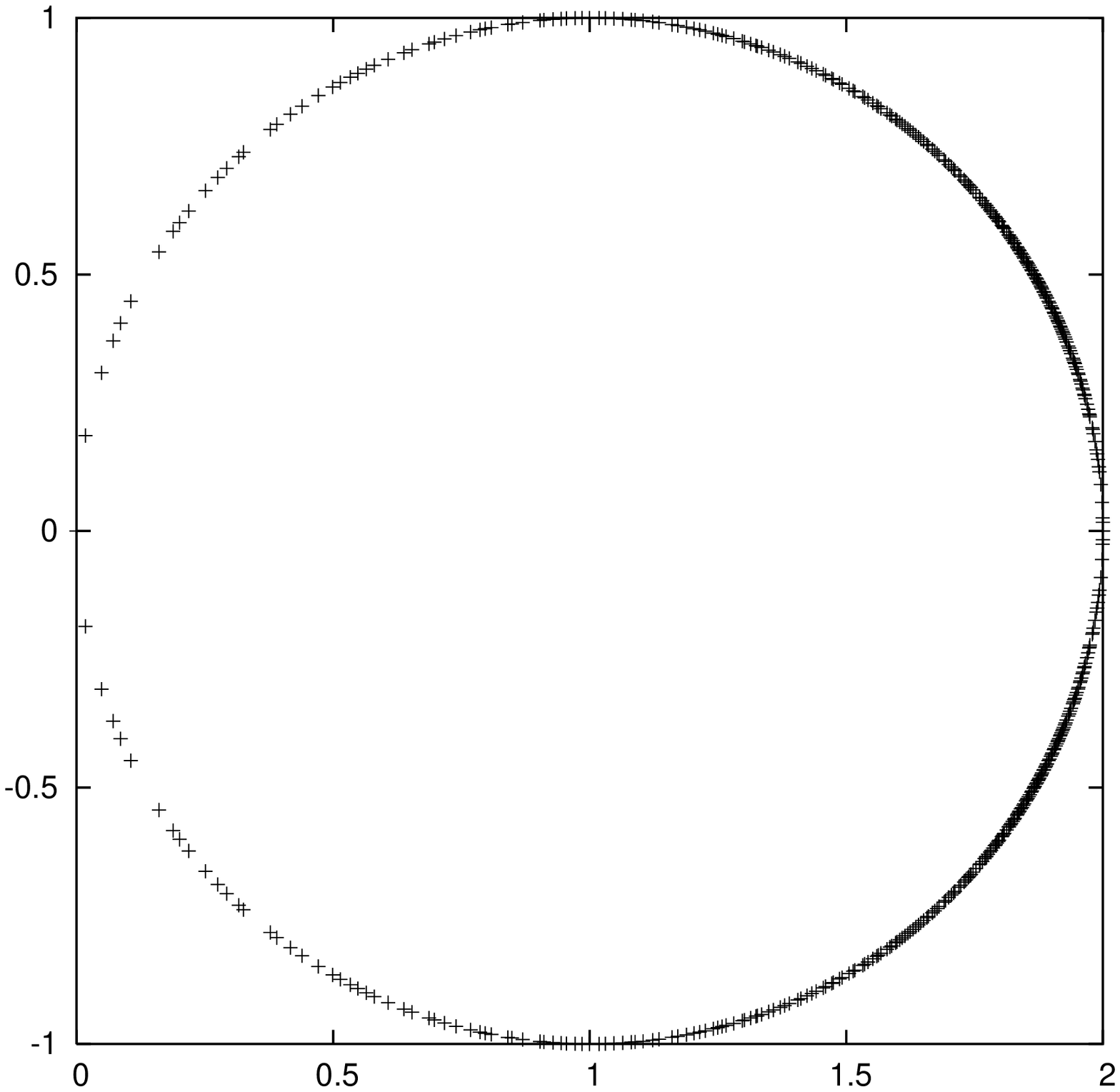}}

\subsection*{Inversion of the Overlap Operator}

Having computed $D$ the next step is its inversion. Before discussing this,
we note that $D^*D$ is optimally inverted using the GGNE algorithm, the reason
being the normality of $D$, i.e. $D^*D=DD^*$. Hence, for dynamical fermion
simulations, which require $D^*D$ inversions, CGNE is the prefered method.
For propagator calculations the situation is less clear. This has to do with
differing spectral properties of $D^*D$ and $D$.

\subsection*{SUMR Algorithm}

We know that GMRES is the optimal method for non-Hermitian operators.
We know also that GMRES requires very often prohibitive computer resources.
However, for unitary matrices one can do better. Exploiting this fact,
one can construct the SUMR or the Shifted Unitary Minimal Residual algorithm
\cite{SUMR_1994},
which is charaterised by short-recurrences and at the same time benefit from
the optimal properties of GMRES:
\begin{small}
\begin{lstlisting}
[rr,x] = SUMR(b,V,rho,zeta,tol,imax);
\end{lstlisting}
\end{small}

\subsection*{Semiconjugate Gradients Algorithm}

We know that the FOM is the counterpart of GMRES for the
Galerkin approach to linear system solvers.
Likewise one can construct the counterpart of
SUMR for shifted unitary systems. This can be done using a new Arnoldi
process for unitary matrices, the Arnoldi Unitary Process. In the
coupled recurrences variant, the search directions of the algorithm
are {\it semiconjugate}. Therefore, the algorithm is called
the Semiconjugate Gradients (SCG) algorithm \cite{SCG}:
\begin{small}
\begin{lstlisting}
[x,rr] = SCG(A,b,x0,tol,nmax);
\end{lstlisting}
\end{small}

\subsection*{Computing Projects}

In this section we compare CGNE, SUMR and SCG algorithms in the same
background U(1) gauge field as before and the same Wilson-Dirac matrix
{\tt A3}:

\begin{lstlisting}
m=0.0001;
D=(1+m)/2*eye(512)+(1-m)/2*V;
b=zeros(512,1);b(1)=1;
[x,rr] = CGNE(D,b,zeros(512,1),1e-12,200);
semilogy(1:2:2*max(size(rr)),rr,'-;CGNE;')
hold
[x,rr] = SCG(D,b,zeros(512,1),1e-12,200);
semilogy(rr,'-.;SCG;')
[rr,x] = SUMR(b,V,(1+m)/2,(1-m)/2,1e-12,200);
semilogy(rr,':;SUMR;')
xlabel("# matrix-vector")
ylabel("residual norm")
replot
gset terminal postscript
gset out "gw_conv_hist.ps"
replot
gset terminal x11
hold
\end{lstlisting}

The result of this comparison is shown in the following figure.
One observes the optimal properties of SUMR, as expected. We note that
SCG is doing worse at the begining until it reaches the
asymtotic regime of SUMR. Both SUMR and SCG are 25\% faster then CGNE.

\centerline{\includegraphics[scale=.6,angle=-90]{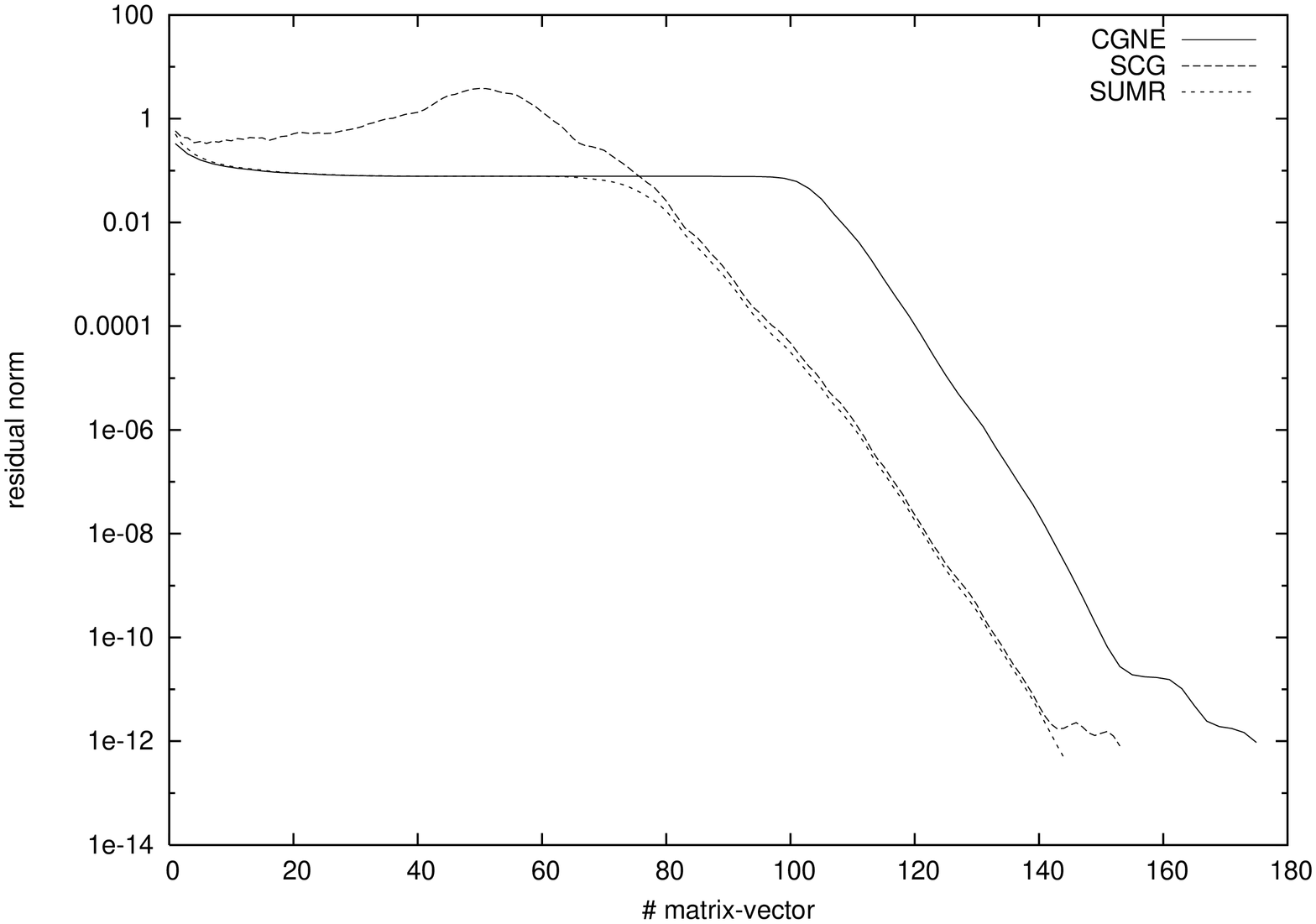}}

\section{QCDLAB 2.x}

In the previous sections we described version 1.0 of QCDLAB. Its
simulation functionality is limited to the QED2 on the lattice,
a very good laboratory for algorithmic and ideas exploration in
lattice QCD. 

We plan to extend functionality of QCDLAB 1.0 further. We are in the
designing phase of QCDLAB 2.0 which will be totaly devoted to lattice
QCD simulation. The gross features of this future version are expected
to be:
\begin{itemize}
\item Dynamically linked functions to already exsisting procedures in
other languages.
\item Ability to compile MATLAB/OCTAVE codes using the Startego Octave
Compiler, {\tt Octave-Compiler.org}.
\item Scalability on various computing platforms using the above compiler.
\item Extended functionality, in particular simulation functions for
lattice QCD and matrix-vector multiplication procedures for various fermion
operators.
\end{itemize}

\end{document}